\documentclass[12pt, preprint]{aastex}

\shorttitle{Atmospheric Dispersion at N-band}
\shortauthors{Skemer et al.}

\begin{document}

\title{A Direct Measurement of Atmospheric Dispersion in N-band Spectra: 
Implications for Mid-IR Systems on ELTs\footnote{The observations reported here were obtained at the MMT Observatory, a facility operated jointly by the Smithsonian Institution and the University of Arizona.}}

\author{Andrew J. Skemer, Philip M. Hinz, William F. Hoffmann, Laird M. Close}
\affil{Steward Observatory, Department of Astronomy, University of Arizona, Tucson, AZ 85721}

\author{Sarah Kendrew, Richard J. Mathar, Remko Stuik}
\affil{Leiden Observatory, Leiden University, The Netherlands}

\author{Thomas P. Greene}
\affil{NASA Ames Research Center, Moffett Field, CA 94035}

\author{Charles E. Woodward}
\affil{Department of Astronomy, School of Physics and Astronomy, University of Minnesota, Minneapolis, MN 55455}

\author{Michael S. Kelley}
\affil{Department of Astronomy, University of Maryland, College Park, MD 20742-2421}

\begin{abstract}
Adaptive optics will almost completely remove the effects of atmospheric turbulence at 10$\micron$ on the Extremely Large Telescope (ELT) generation of telescopes.  In this paper, we observationally confirm that the next most important limitation to image quality is atmospheric dispersion, rather than telescope diffraction.  By using the 6.5 meter MMT with its unique mid-IR adaptive optics system, we measure atmospheric dispersion in the N-band with the newly commissioned spectroscopic mode on MIRAC4-BLINC. Our results indicate that atmospheric dispersion is generally linear in the N-band, although there is some residual curvature.  We compare our measurements to theory, and make predictions for ELT Strehls and image FHWM with and without an atmospheric dispersion corrector (ADC).  We find that for many mid-IR applications, an ADC will be necessary on ELTs.

\end{abstract}

\section{Introduction}
As we approach the Extremely Large Telescope (ELT) generation of telescopes, adaptive optics is becoming increasingly important to the general astronomical community.  Large telescopes have small diffraction limits, and achieving these limits is a major goal for instrument builders.  The mid-infrared wavelengths, in particular, stand to gain substantially from larger telescopes---at the diffraction limit, S/N $\propto$ Diameter$^{2}$ for background-limited observations of point-sources.  Today's $\sim$8-meter class telescopes are close to diffraction-limited in the mid-infrared, even without adaptive optics, but maintaining the diffraction limit as telescopes continue to scale upwards will be challenging.

For ground-based telescopes in the mid-infrared, seeing is often considered a minor effect, and other (smaller) atmospheric effects are completely ignored.  \citet{2008SPIE.7015E.159K} have predicted several atmospheric properties that may limit image quality on ELTs, including mid-infrared atmospheric dispersion\footnote{In this paper, we use ``refraction" to refer to the absolute bending of light and ``dispersion" to refer to the differential chromatic bending of light.  Generally, theoretical considerations use refraction while practically, telescope images are affected by dispersion.}, visible atmospheric dispersion for wavefront sensing, and water vapor turbulence (see \citet{2008ApOpt..47.1072D} for a similar discussion in the near-infrared).  So far, these effects have not been adequately measured. 

The 6.5 meter MMT, with its unique mid-IR adaptive optics system (MMTAO) provides a powerful testbed for mid-IR AO on ELTs.  Effects that will severely limit image quality on ELTs are just measurable with MMTAO due to its highly stable PSF.  By removing the largest atmospheric effect (seeing), we measure the second largest effect (atmospheric dispersion) with the newly commissioned spectroscopic mode of the MMT's Mid-Infrared Array Camera (MIRAC4-BLINC) and adaptive optics.

Atmospheric refraction is a well-known phenomenon at visible wavelengths, where it is typically treated as a smooth curve that flattens quickly longward of K-band \citep{Edlen, Ciddor, Bonsch}.  However, more detailed treatments show that molecular resonances from CO$_{2}$ and H$_{2}$O (amongst others) dominate the infrared refractivity curve \citep{1986InfPh..26..371H, mathar, 2004PASP..116..876C, 2007JOptA...9..470M}.  These authors show that each infrared window (L,M,N) is  bracketed by molecular absorption and has an atmospheric refraction curve  characterized by an S-shape superimposed on a stronger linear trend.  

In this paper, we measure the atmospheric dispersion curve on the short wavelength side of N-band (8.26$\micron$-11.27$\micron$) using spectroscopy and adaptive optics.  In previous studies, \citet{1999PASP..111..512L} measured refractivity at one wavelength (12$\micron$) while  \citet{2004SPIE.5491..588T} interferometrically measured refractivity throughout the N-band but were insensitive to the overall trend.  Our spectroscopic result has the benefit of measuring all wavelengths simultaneously, so that the overall trend and curvature of the effect throughout N-band is unambiguous.  By directly measuring the atmospheric dispersion curve, we can assess how the effect will limit image quality in the mid-infrared for ground-based ELTs.  This is useful for instrument builders, who will have the option of using atmospheric dispersion correctors (ADCs) to suppress the effect. 

\section{Observations and Instrument Description}
     Our data were obtained March 4, 2009 UT with the 6.5 meter MMT and its deformable
secondary adaptive optics system \citep[MMTAO -- e.g.,][]{2000PASP..112..264L, 2003SPIE.5169...17W, 2004SPIE.5490...23B}. We used the newly commissioned spectroscopic mode of MIRAC4-BLINC.  The instrument is a combination of the Mid-IR Array Camera, Gen. 4 (MIRAC4) and the Bracewell Infrared Nulling Cryostat \citep[BLINC --][]{2000SPIE.4006..349H} which for these observations, is used in its ``imaging" mode.  MIRAC4 is functionally similar to previous incarnations of MIRAC \citep[e.g.][]{1998SPIE.3354..647H} with the main new feature being a DRS Technologies 256 x 256 Si:As array. Some of the relevant details of this new instrument are described below. 

MIRAC4-BLINC was use to observe the mid-infrared standards $\alpha$ Her and  $\gamma$ Aql. Both targets were bright in the visible (3.06 and 2.72 V magnitudes for $\alpha$ Her and $\gamma$ Aql, respectively) which allowed us to run the MMTAO system at full sampling speed
(550 Hz). At longer wavelengths the MMTAO system can produce nearly perfect diffraction-limited images with extremely stable point spread functions \citep{2004SPIE.5490..351K, 2006ApJ...653.1486H}.  At N-band, typical Strehls of up to $\sim$98\% can be obtained under good seeing conditions \citep{2003ApJ...598L..35C}. Conditions were non-photometric with moderately high winds (bursts up to 30 mph).  However, the adaptive optics system was consistently able to stay locked on bright sources.  Data from
the MMT weather station showed an average temperature of 7.8 $^{\circ}$C, an average pressure of
745 mbar and an average relative humidity of 44.3\%. Detailed weather descriptions for each observation are shown in Table \ref{Observations}.

The MIRAC4-BLINC optics are enclosed and cooled in two attached cryostats.  Reflective reimaging optics in the BLINC portion of the system create an image of the secondary on an articulated mirror. The mirror provides rapid chopping capability at rates of 1-10 Hz.  A mask overlaid on the mirror provides the critical cold stop for the system.  Downstream from the chopping mirror an image of the telescope focal plane is formed at an f-ratio of f/26.7 (1.22 arcsec/mm).  A cold image stop wheel allows insertion of several slits at this location.  A second set of reflective reimaging optics within the MIRAC4 cryostat create a second pupil image at a cold stop, which is followed by two filter wheels, the first of which contains a KRS-5 grism.  The grism was fabricated by Zeiss and has an 11 degree wedge and 27 lines /mm to create a first-order, low-resolution (R$\sim$100) spectrum on the detector.  Flexure of the grism, with respect to the applications in this paper, is negligable.  By adjusting the position of the detector as well as a set of fold optics it is possible to create a range of magnifications from 0.55-1.1 (giving final plate scales from 0.054 to 0.11 "/pixel, for the 50 micron pixels of the MIRAC4 detector).  

For these observations we used the high magnification setting of the camera to maximize our PSF sampling.  At maximum magnification the grism dispersion is 12.6 nm/pixel.   We used an off-center, 0.6 mm wide (0.73") slit that placed the spectrum from 8.26-11.27 microns on the array.  At 10 microns wavelength this gives a diffraction-limited spectral resolution of 130.  The dispersion and range of the spectra were calibrated by using a thin transmissive piece of polystyrene and correlating the known spectrum of the film with the measured response of the system.

\begin{figure}
 \includegraphics[angle=0,width=\columnwidth]{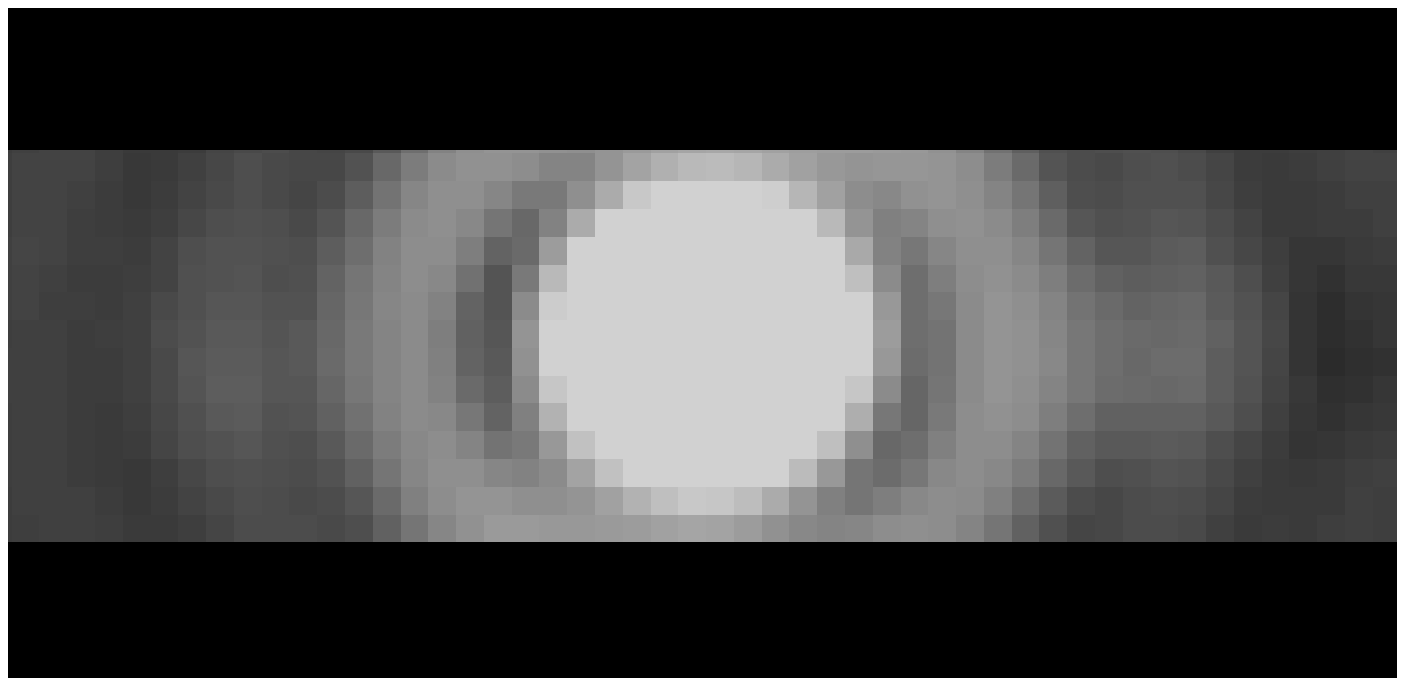}
\caption{Image of an MMTAO 10.55$\micron$ PSF (Pollux) with our 0.75" slit superimposed.  The high-Strehl core is completely contained in the slit.  \citep[Pollux PSF image from][]{2008ApJ...676.1082S}
\label{slit_image}}
\end{figure}

     We aligned the target in the slit for both nod beams (see Figure \ref{slit_image}),
chopping perpendicular to the slit and nodding along the slit. The slit was aligned perpendicular to the horizon in order to optimally measure atmospheric dispersion.
Details of our observations are listed in Table \ref{Observations}.

\begin{figure}
 \includegraphics[angle=90,width=\columnwidth]{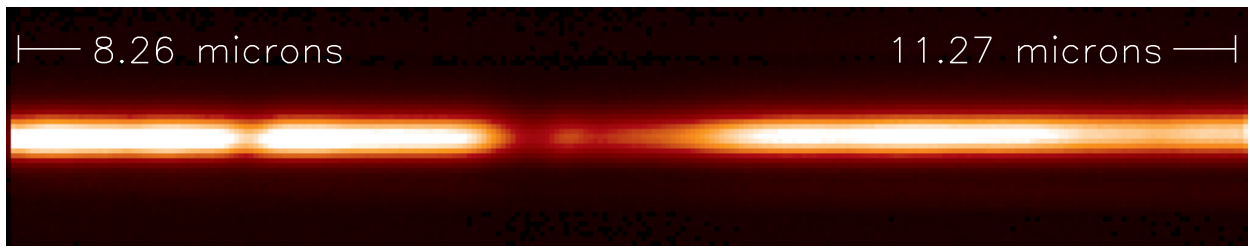}
\caption{The raw, combined image of $\alpha$ Her's spectrum at 1.05 airmasses taken with MIRAC4-BLINC and MMTAO.  Because the
data were taken in MIRAC4's high-magnification mode, our PSF is well-sampled with 0.055" pixels (compared to the $\sim$0.3" MMTAO
diffraction-limited FWHM).  The combination of high Strehl and a fine platescale allow us to measure the trace of the spectrum (centroid at each wavelength) to high precision.
\label{spectrum_image}}
\end{figure}

     The data were reduced using our custom artifact removal software described in \citet{2008ApJ...676.1082S}. The images were then cross-correlated (with spline interpolation) and median combined (see a reduced image in Figure \ref{spectrum_image}).

\section{Analysis}
\begin{figure}
 \includegraphics[angle=0,width=\columnwidth]{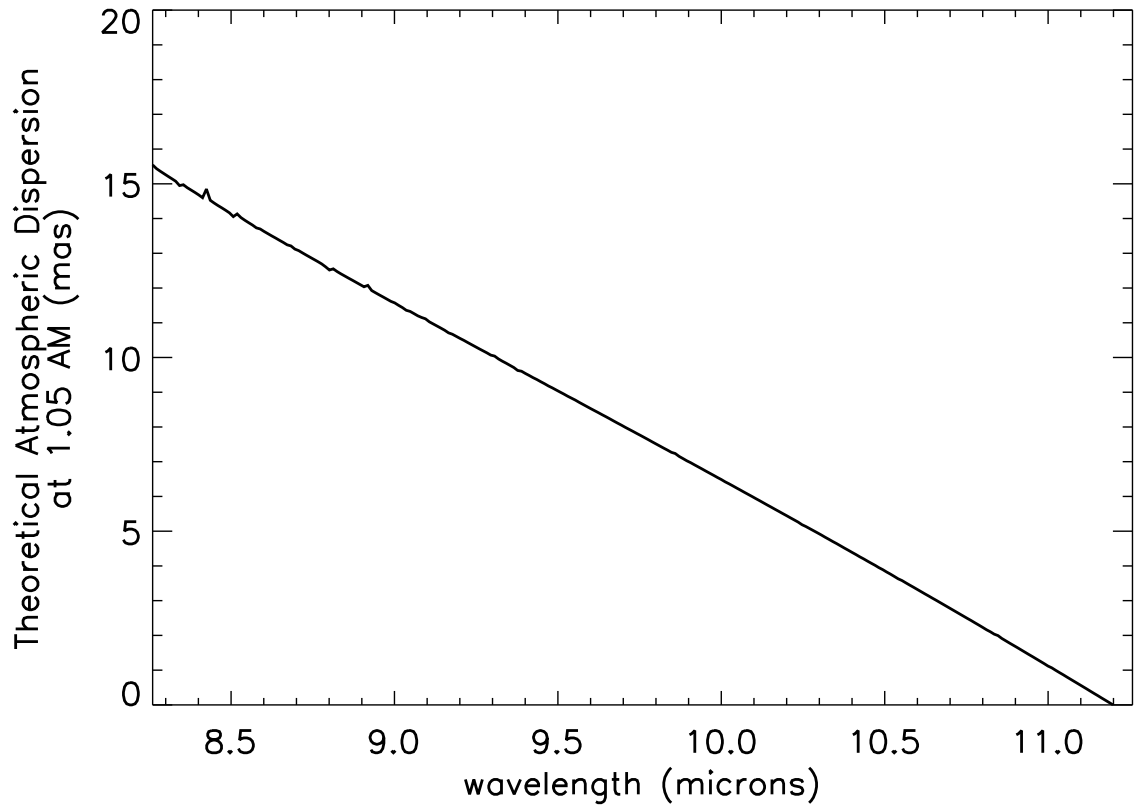}
\caption{Theoretical atmospheric dispersion across our spectroscopic band at 1.05 airmasses from the models of \citet{mathar, 2007JOptA...9..470M}.  Our measurements subtract the atmospheric dispersion at 1.05 airmasses so that we can measure our grism's intrinsic alignment/curvature.  This means our measurements consistently underestimate atmospheric dispersion by the amount shown in this plot ($\sim$0.015").
\label{1.05dispersion}}
\end{figure}

In order to determine the intrinsic curvature of the MIRAC4 grism, we assume the $\alpha$ Her data taken at 1.05 airmasses is unaffected by atmospheric dispersion.  The models of \citet{mathar, 2007JOptA...9..470M} suggest that the effect is small, but non-negligable (see Figure \ref{1.05dispersion}).  As a result, all of our measurements underestimate the effect of atmospheric dispersion by the amount shown in Figure \ref{1.05dispersion} at each airmass (a cumulative effect of 0.015").  In the interest of not contaminating our measurements with models, we ignore the effect for the rest of the analysis section.  However, we do include the extra atmospheric dispersion in our ELT implications section.

\begin{figure}
 \includegraphics[angle=90,width=\columnwidth]{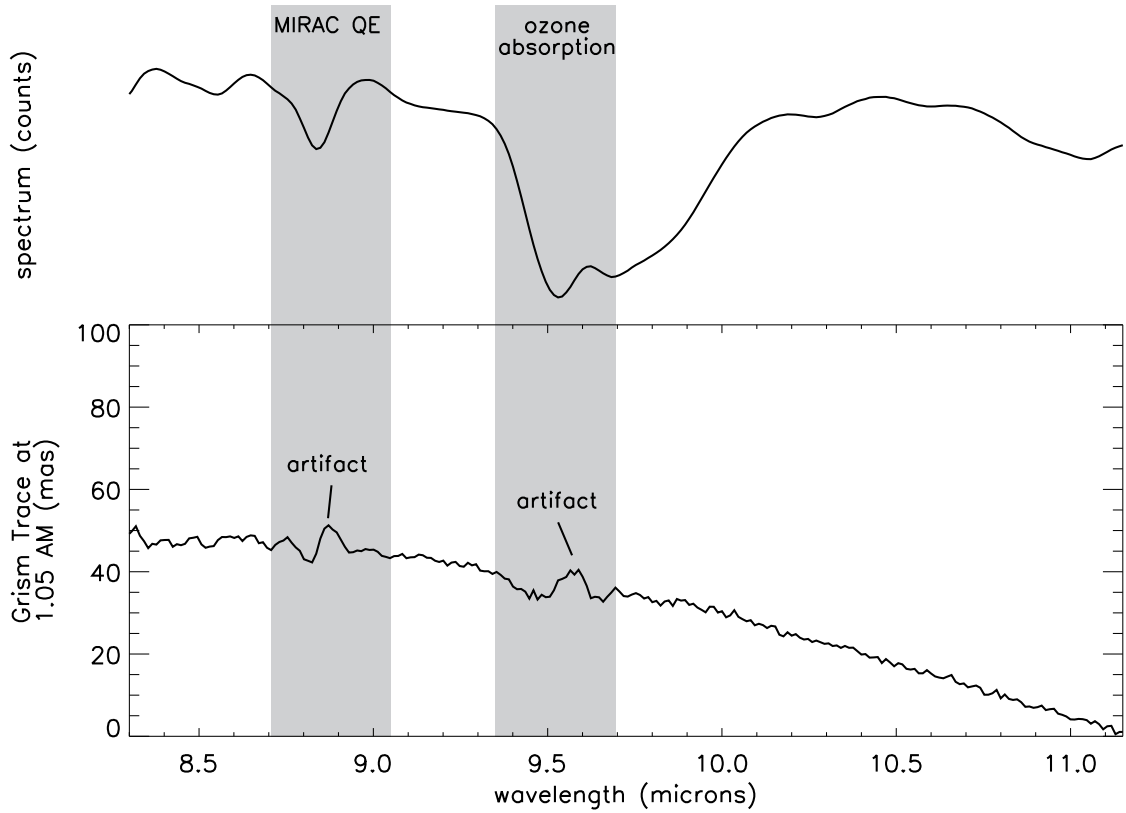}
\caption{TOP: N-band spectrum of $\alpha$ Her taken at 1.05 airmasses.  The spectrum has not been flux calibrated, and shows two large features:  at $\sim$8.8$\micron$, there is a detector quantum efficiency drop and at $\sim$9.7$\micron$ there is telluric ozone absorption.  BOTTOM: Grism spectral trace of $\alpha$ Her at 1.05 airmasses (where atmospheric dispersion is assumed to be low; see Figure \ref{1.05dispersion}). The trace features coincident with the MIRAC quantum efficiency (QE) effect and telluric ozone are artifacts. 
\label{grism_trace}}
\end{figure}

We measure the trace of the grism by centroiding every wavelength with a best-fit, Moffat function \citep{2009arXiv0902.2850M}.  Four parameters (peak value, centroid, FWHM and Moffat Index) were allowed to vary with wavelength, with centroid being the relevant parameter.  In principle, this fit could be used to produce an error estimate derived from random (ie photon) noise.  However, our dominant error source is systematic noise caused by psf-mismatch (described below).  The grism trace is shown in Figure \ref{grism_trace} and shows a 0.05" offset over the range of our spectrum.  The linear trend is the result of grism alignment and the curvature of the trace is intrinsic to the optical system.  Two dips appear in our raw spectrum, which correspond to apparent artifacts in the grism trace.  Since our PSF is a spectrally smoothed two-dimensional image, spectral features can easily cause slight PSF mismatches, which manifest themselves as grism trace artifacts.  These regions are ignored in any further analysis.

\begin{figure}
 \includegraphics[angle=0,width=\columnwidth]{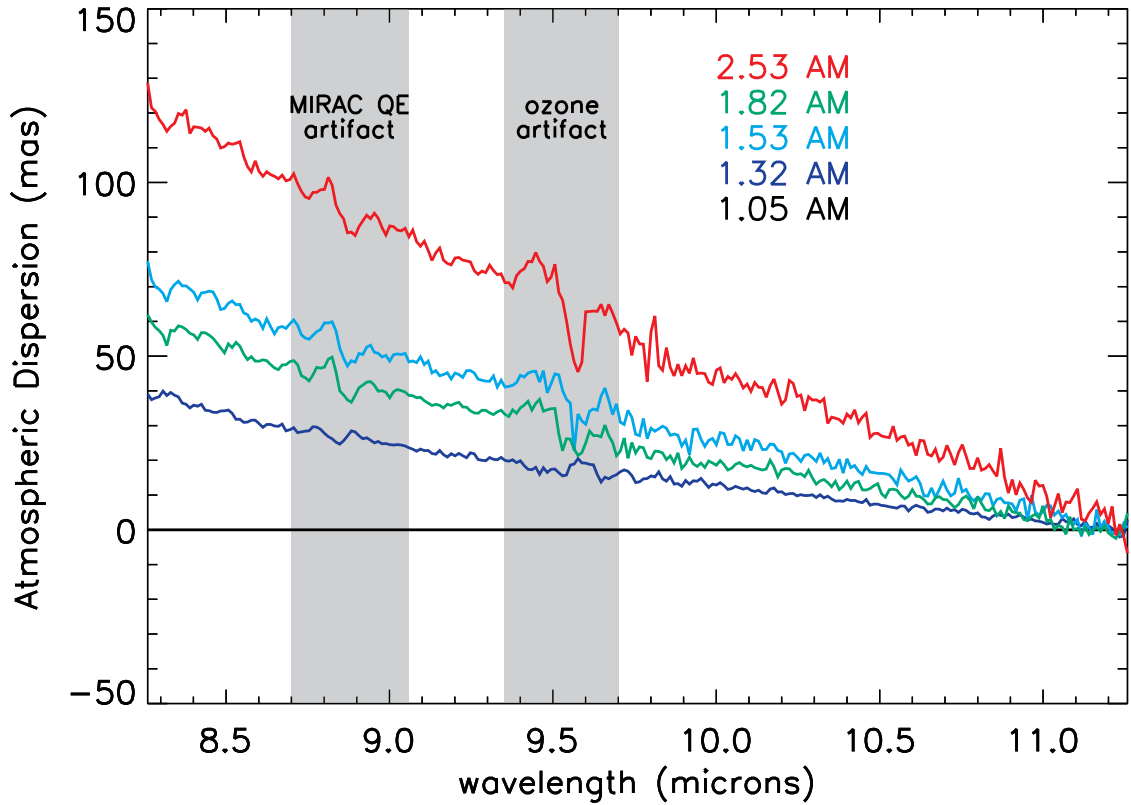}
\caption{MIRAC4-BLINC/MMTAO measurements of atmospheric dispersion in the N-band.  For each airmass, we measure the grism trace and subtract the grism's intrinsic curvature (Figure \ref{grism_trace}).  The effect has been fixed to 0 at $11.2\micron$.  Our results show that atmospheric dispersion in the mid-infrared is a relatively large effect, although it is considerably smaller than the $\sim$0.3" diffraction-limited FWHM of a 6.5 meter telescope.  On larger ELTs, mid-IR atmospheric dispersion will severely limit image quality if left uncorrected.  Note that we underestimate atmospheric dispersion by the amount shown in Figure \ref{1.05dispersion} due to our assumption that atmospheric dispersion is negligible at 1.05 airmasses.
\label{refraction}}
\end{figure}

For each observation listed in Table \ref{Observations}, we repeat the trace measurement, and subtract the grism's intrinsic curvature (the trace of the 1.05 airmass spectrum; Figure \ref{grism_trace}).  This gives us a direct measurement of the atmospheric dispersion across our spectral range (minus the atmospheric dispersion at 1.05 airmasses).  The effect is fixed to 0 at the red end (11.2 $\micron$) of our spectrum and is shown for each observed airmass in Figure \ref{refraction}.  There is clear evidence that the blue light is refracted more than the red, and that this trend increases with airmass.  A fit of the linear trend between 9.9 and 11.0 $\micron$ gives 10.3 mas/$\micron$, 20.4 mas/$\micron$, 15.8 mas/$\micron$ and 33.9 mas/$\micron$ at 1.32 airmasses, 1.53 airmasses, 1.82 airmasses and 2.53 airmasses respectively.  We note that the observed trend is not perfectly sequential as our 1.53 airmass data appear to have experienced more atmospheric dispersion than our 1.82 airmass data.

\begin{figure}
 \includegraphics[angle=0,width=\columnwidth]{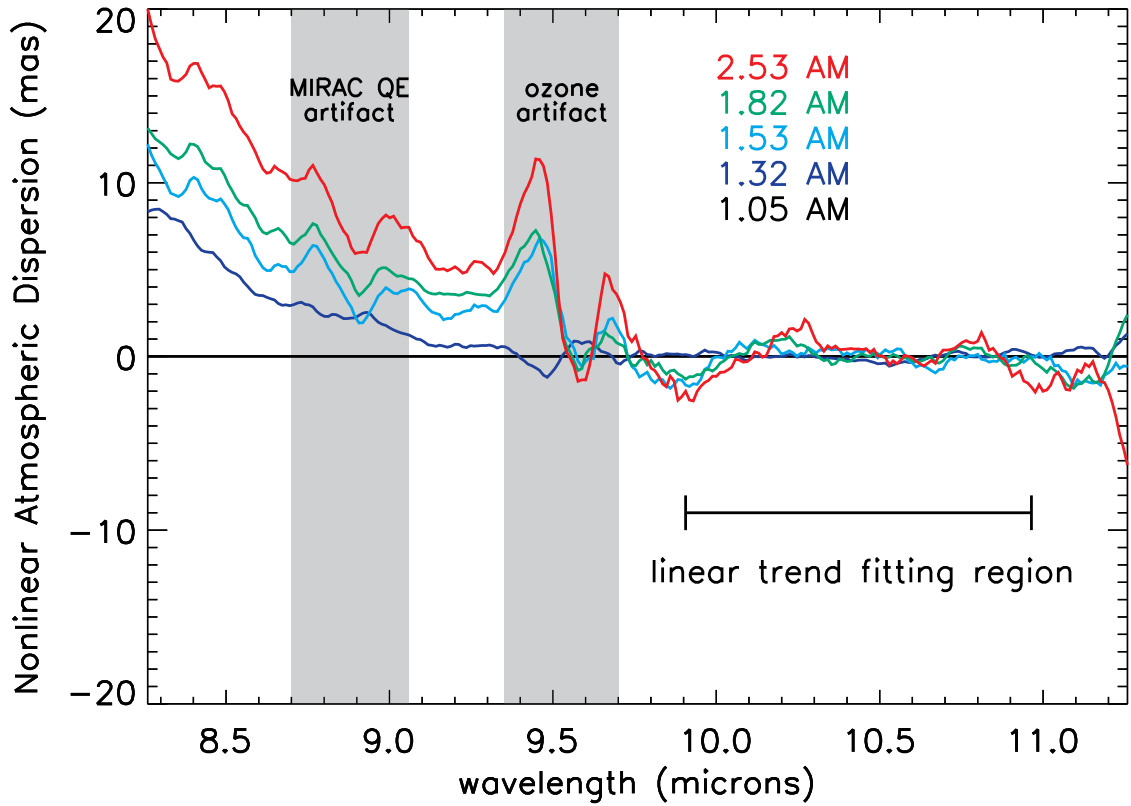}
\caption{On ELTs, mid-IR atmospheric dispersion correctors (ADCs) will be necessary to suppress the linear atmospheric refraction effect shown in Figure \ref{refraction}.  In this figure we fit and remove the linear trend from each atmospheric dispersion curve.  The resultant nonlinear residuals will be uncorrected by traditional ADCs.
\label{refraction_curvature}}
\end{figure}

The strong linear trends from Figure \ref{refraction} imply that atmospheric dispersion is an important effect to consider when designing mid-IR instruments on large telescopes.  For the next (ELT) generation of telescopes, a mid-IR ADC will be necessary to achieve diffraction-limited images.  However, traditional ADCs can only correct linear atmospheric dispersion.  In Figure \ref{refraction_curvature} we show the \textit{curvature} of atmospheric dispersion by subtracting off a linear trend from the data shown in Figure \ref{refraction}.  This simulates the effect of atmospheric dispersion after correction from a perfectly tuned ADC.  Unfortunately, some elongation of the PSF may still occur in broad filters even with an ADC due to nonlinear atmospheric dispersion.  Note that the observed curvature is sequential with airmass.

\section{Discussion}

\subsection{Fitting with Models}

We compare our measurements to the models described by \citet{mathar, 2007JOptA...9..470M}.  These models calculate refractive index values ($n$) by summing over the electronic transitions of atmospheric molecular constituents from the far-ultraviolet to the far-infrared, using the molecular line database HITRAN \citep{2005JQSRT..96..139R}.  A full description of the model can be found in \citet{mathar}.

The refraction, or the angular distance between the true and apparent zenith distances for a given refractive index, can be calculated from:
\begin{equation}\label{eq:R}
R  \approx 206,265(\frac{n^2-1}{2n^2})\tan(z)
\end{equation}
where $z$ is the true zenith distance in radians and $R$ is in arcseconds. From
this, the differential refraction (dispersion) between two wavelengths is given
by:
\begin{equation}\label{eq:adr}
R_1-R_2 \approx 206,265(\frac{n_1^2-1}{2n_1^2}-\frac{n_2^2-1}{2n_2^2})\tan(z).
\end{equation}

\begin{figure}
 \includegraphics[angle=90,width=\columnwidth]{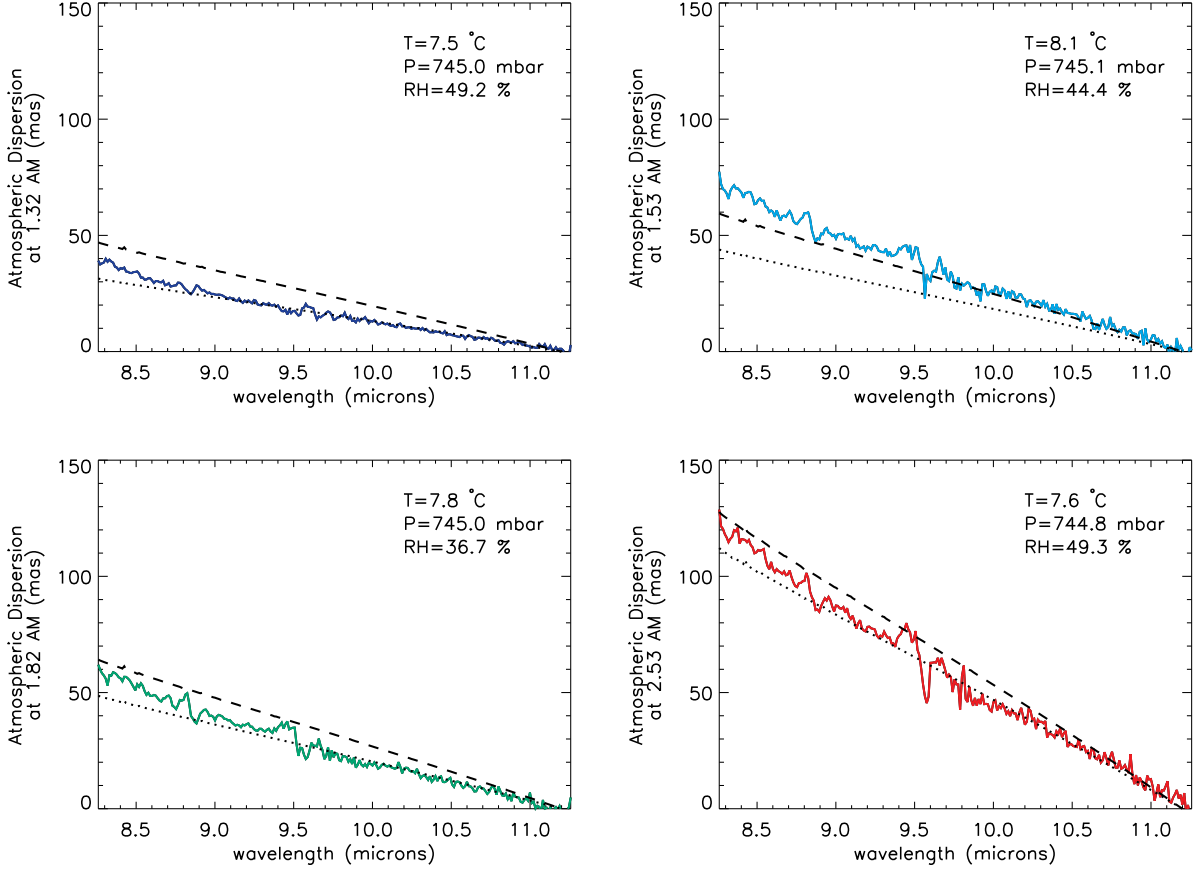}
\caption{Comparisons of our measured atmospheric dispersion with theory from \citet{mathar, 2007JOptA...9..470M}.  Our measurements consistently underestimate atmospheric dispersion by the amount shown in Figure \ref{1.05dispersion}, due to our differential measurement (with a 1.05 airmass grism trace).  In all four plots, the solid, colored curves are our measurements, the dotted curves are models that subtract the 1.05 airmass dispersion from Figure \ref{1.05dispersion} to properly account for the underestimate described above, and the dashed curves are the true models.  The models all have been calculated using weather data from the corresponding observation (see Table \ref{Observations}, and have errors of about 10\% based on the varying humidity during each observation.  Three of the four models are very good fits to the linear trend of our measurements.  However, the models all indicate less curvature than is seen in the data.
\label{modelcompare}}
\end{figure}

We calculate the expected atmospheric dispersion using the output of Mathar's models and equation \ref{eq:adr} (where the value of $n_2$
is fixed to the refractive index at $\lambda =11.2\micron$, as we did in our differential MMTAO measurements).  Because our measurements subtract off atmospheric dispersion at 1.05 airmasses (assuming it to be small; see Figure \ref{1.05dispersion}), we do the same with our models.  The comparison between our measurements and the models is shown in Figure \ref{modelcompare}.  The solid curves are our measurements, the dotted curves are the models with dispersion at 1.05 airmasses subtracted and the dashed curves are the models without dispersion at 1.05 airmasses subtracted.  

We ran the models at a variety of relative humidities to reflect the measured variation during each group of observations (see Table \ref{Observations}).  These small humidity variations create a $\sim$10\% model uncertainty for each set of observations.

The models show a good fit to the observed spectral trace longward of $\sim$9.5
$\micron$ in three of the four cases (qualitatively, a quantitative approach, such as a $\chi^{2}$ test would be inappropriate given that our dominating errors are systematic).  The clear
exception is the spectral trace taken at airmass 1.53, which follows the same
trend as the others but lies significantly above the curve predicted for the
meteorological conditions at the time of observation. The discrepancy could be due to a temporary burst of moisture high up in the atmosphere (where our ground-based weather monitors are insensitive).  It could alternatively be the result of a filter wheel return error which caused our grism to be slightly misaligned.  This second scenario is unlikely as the filter wheel has a resolution of $<$0.03$^{\circ}$, which should provide a spectrum tilt of, at most, 5 mas across the array.

At all four airmasses, our measurements show more curvature than the models predict.  At this point, it is unclear whether the source of disagreement is error in the measurements or error in the models.  Given that the linear trend dominates the curvature, this disagreement may prove insignificant.  However, if the linear trend is corrected with an ADC, the curvature may still cause some non-negligble dispersion at very high airmasses.

Overall, the Mathar models fit our measurements very well, and will be useful for ADC designs and operations.  It is still an open question whether ground-based weather measurements will make good predictors of atmospheric dispersion throughout the atmosphere.  In our case ground-based weather measurements allowed accurate predictions in three out of four cases.  The source of the residual curvature is also unclear at this point.  Both issues should be addressed with future observations.

\subsection{Implications for ELTs}

Mid-IR cameras on ELTs will need to operate in several modes to accomplish a wide variety of scientific tasks.  The loss of image quality related to atmospheric dispersion will affect each situation differently.  Here we discuss specific implications to three commonly used mid-IR modes: broad-band imaging, narrow-band imaging and spectroscopy.

\subsubsection{Broad-Band Imaging}

Using our MMTAO observations, we can simulate the degradation of image quality for ELTs.  Our MMTAO observations only cover 8.26$\micron$-11.27$\micron$ but the broad N-band extends all the way to $\sim$14$\micron$.  We approximate full N-band dispersion curves (8.26$\micron$-13.74$\micron$) by reflecting our MMTAO dispersion curves about 10.5$\micron$ (creating the characteristic S-shape; this may be an oversimplification as \citet{mathar, 2007JOptA...9..470M}'s models show slightly increased curvature longward of 11$\micron$).  We also add back the theoretical dispersion at 1.05 airmasses having confirmed the validity of the linear trends in \citet{mathar, 2007JOptA...9..470M}'s models.  Finally, we fit the curves with a fifth-order polynomial to remove the noise and systematics shown in Figures \ref{grism_trace}-\ref{refraction_curvature}.

\begin{figure}
 \includegraphics[angle=0,width=\columnwidth]{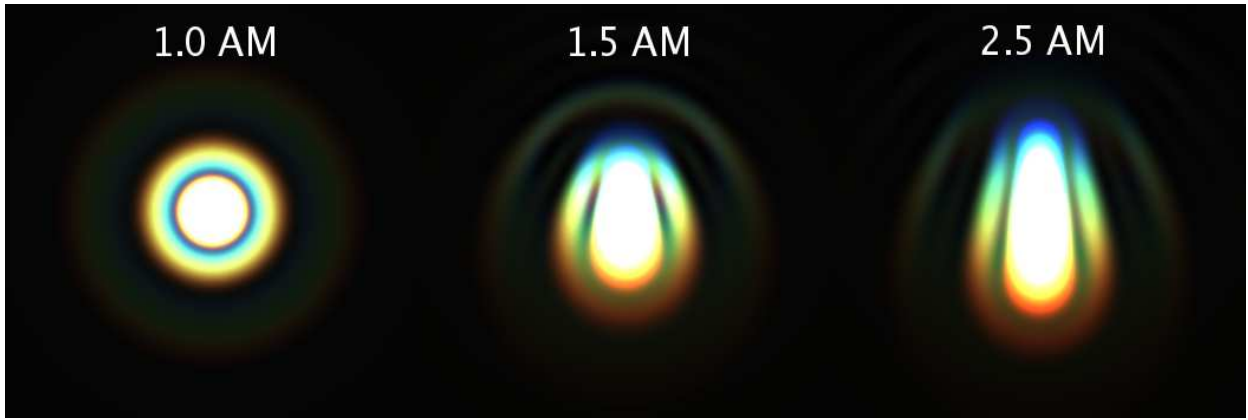}
\caption{Simulated 3-color N-band images (blue is 8.26$\micron$ and red is 13.74$\micron$) for a 42 meter telescope at different airmasses (zenith is up).  Without an ADC, image quality will be significantly degraded in the altitude axis.  The images assume a flat SED and a site similar to the MMTs.
\label{E_ELT_PSF}}
\end{figure}

We simulate ELT images by convolving our estimated dispersion curves with diffraction-limited PSFs and flat SEDs.  In this case, the (1-D) convolution step is simply adding together PSFs at positions defined by our dispersion curves.  The PSFs are constructed from annulus apertures with outer diameters 24.5$m$, 30$m$ and 42$m$ and a 20\% central obscuration.  An example of our simulated images is shown for a 42 meter telescope with no ADC in Figure \ref{E_ELT_PSF}.  The results show a severe elongation in the altitude axis.

Using our simulated ELT images, we measure Strehl and FWHM, with and without a linear ADC, at 1.0, 1.5 and 2.5 airmasses.  These results are shown in Table \ref{ELT_predictsN}.  Without a linear ADC, Strehl and FWHM are significantly degraded for all three telescopes, even at 1.5 airmasses.  With a linear ADC, the images are almost perfectly corrected at reasonable airmasses.  Thus, we conclude that a linear ADC is essential for broad N-band imaging on ELTs and that a higher order, ``non-linear" ADC is not.

\subsubsection{Narrow-Band Imaging}

We repeat the experiment from the previous section but with a 10\% filter centered at 10.5$\micron$ (note that no reflection of the atmospheric dispersion curve is necessary).  Our results are shown in Table \ref{ELT_predictsnarrow}.  Image quality is still noticeably degraded with the narrow-band filter, but the effect only becomes serious for the largest (42$m$) telescope.  With an ADC, narrow-band imaging is completely unaffected by atmospheric dispersion.  Thus, we conclude that a linear ADC will be useful for some narrow-band imaging applications on ELTs, and unnecessary for others based on scientific needs.

\subsubsection{Spectroscopy}
The chromaticity of the images shown in Figure \ref{E_ELT_PSF} imply that chromatic slit loss may be a major problem for ELT mid-IR spectroscopy.  The problem can be avoided by always keeping the slit perpendicular to the horizon so that there is no refraction in the spectral dispersion axis.  However, this would preclude observations of a variety of spatially resolved objects (binaries, circumstellar disks, etc.) where a properly aligned position angle is important.  Based on our predictions for broad-band image quality (Table \ref{ELT_predictsN}), a linear ADC would suppress dispersion to the point where broad-band spectroscopy would be possible at different position angles, given a wide slit.  For certain high resolution spectroscopic applications (with a very narrow bandpass), an ADC will not be necessary.

\section{Conclusions}
After seeing is removed by adaptive optics, atmospheric dispersion will be the dominant source of image quality degradation on ground-based ELTs, surpassing diffraction.  While theory has predicted large S-shaped refraction curves in each infrared window, the effect had not been measured as a function of wavelength through the atmosphere.  In this paper, we use MMTAO and the MIRAC4-BLINC spectrograph to measure atmospheric dispersion from 8.26$\micron$-11.27$\micron$.  We find the following:

1) ``Blue"-light (8.26$\micron$) is refracted more than ``red"-light (11.27$\micron$) in the mid-infrared, and the effect increases with airmass and humidity.

2) Although the linear term of mid-IR atmospheric dispersion is dominant, a non-negligable amount of curvature exists too.  Observations over different wavelength ranges (including full N-band) will help determine the source of this curvature.

3) The dominating linear trends in our measurements are in excellent agreement with the models of \citet{mathar, 2007JOptA...9..470M}.  We measure more curvature than the theory predicts; however, the magnitude of the curvature is small compared to the linear trend.  \citet{mathar, 2007JOptA...9..470M} models may now be used to develop predictive models for ADCs given ground-based measurements of temperature, pressure, relative humidity and airmass.

4) Based on simulations of mid-IR ELT adaptive optics images with atmospheric dispersion, we find that ADCs will be useful for high-Strehl, narrow-band imaging and spectroscopy, and essential for high-Strehl, broad-band imaging and spectroscopy.  Our conclusions are only based on an analysis of image quality.  We make no claims about the technical feasibility (cost, increased background, decreased throughput, etc.) of a mid-IR ADC.  Instrument builders will have to weigh these issues as well.

\acknowledgements
We wish to thank Timothy Pickering for his help acquiring MMT weather data and Jared Males for providing the Airy pattern generating code that was used in our simulations.  We also thank Matthew Kenworthy and Derek Kopon for useful conversations and feedback.  AJS acknowledges the NASA Graduate Student Research Program (GSRP) and the University of Arizona's Technology Research Initiative Fund (TRIF) for their generous support.  LMC is supported by an NSF MRI and TSIP award. CEW also acknowledges support from the National Science Foundation grant AST-0706980.

\clearpage
\begin{deluxetable}{lcccccccccccc}
\tabletypesize{\scriptsize}
\tablecaption{MMTAO Observations and Weather (March 4, 2009 UT)}
\tablewidth{0pt}
\tablehead{
\colhead{Object} &
\colhead{Airmass} &
\colhead{Exposure Length} &
\colhead{\# of Exposures} &
\colhead{Pressure} &
\colhead{Temperature} &
\colhead{Relative Humidity}
\\
\colhead{} &
\colhead{} &
\colhead{(s)} &
\colhead{} &
\colhead{(mbar)} &
\colhead{($^{\circ}$C)} &
\colhead{(\%)}
}

\startdata

Alpha Her & 1.05 & 0.05 & 80 & 745.1 & 8.0 & 42.1$\pm$0.5 \\
Alpha Her & 1.32 & 0.05 & 80 & 745.0 & 7.5 & 49.2$\pm$1.5 \\
Gamma Aql & 1.53 & 3 & 16 & 745.1 & 8.1 & 44.4$\pm$0.4 \\
Gamma Aql & 1.82 & 3 & 26 & 745.0 & 7.8 & 36.7$\pm$1.3 \\
Gamma Aql & 2.53 & 3 & 16 & 744.8 & 7.6 & 49.3$\pm$0.5 \\

\enddata
\tablecomments{Values listed for airmass, pressure, temperature and relative humidity are averages over each group of observations, with the error bars on relative humidity showing the standard deviation of the range of measured values over each group of observations. The weather data were recorded outside the telescope dome using a standard weather monitoring station on a 20 foot pole.  All data were taken with an AO loop-speed of 550 Hz.}
\label{Observations}
\end{deluxetable}

\clearpage

\begin{deluxetable}{ccccccccccccc}
\tabletypesize{\scriptsize}
\tablecaption{Predicted N-band Image Quality for ELTs}
\tablewidth{0pt}
\tablehead{
\colhead{Telescope Diameter (m)} &
\colhead{Airmass} &
\colhead{Strehl (\%)} &
\colhead{Strehl with ADC (\%)} &
\colhead{FWHM (mas)} &
\colhead{FWHM with ADC (mas)}
}

\startdata
42   & 1.0 & 100 & 100 &  53 & 53 \\
42   & 1.5 &  43 &  98 & 135 & 54 \\
42   & 2.5 &  27 &  94 & 226 & 56 \\
\hline
30   & 1.0 & 100 & 100 &  74 & 74 \\
30   & 1.5 &  56 &  99 & 139 & 75 \\
30   & 2.5 &  37 &  97 & 225 & 77 \\
\hline
24.5 & 1.0 & 100 & 100 &  91 & 91 \\
24.5 & 1.5 &  64 &  99 & 145 & 92 \\
24.5 & 2.5 &  44 &  98 & 225 & 93 \\

\enddata
\tablecomments{The N-band filter is assumed to be rectangular from 8.26$\micron$-13.74$\micron$.  We also assume a flat SED and a site similar to the MMT's.  FWHM is measured in the altitude axis.  In the azimuth axis, FWHM is assumed to be diffraction-limited.}
\label{ELT_predictsN}
\end{deluxetable}

\clearpage

\begin{deluxetable}{ccccccccccccc}
\tabletypesize{\scriptsize}
\tablecaption{Predicted 10\%-band Image Quality for ELTs}
\tablewidth{0pt}
\tablehead{
\colhead{Telescope Diameter (m)} &
\colhead{Airmass} &
\colhead{Strehl (\%)} &
\colhead{Strehl with ADC (\%)} &
\colhead{FWHM (mas)} &
\colhead{FWHM with ADC (mas)}
}

\startdata
42   & 1.0 & 100 & 100 & 52 & 52 \\
42   & 1.5 &  94 & 100 & 55 & 52 \\
42   & 2.5 &  88 & 100 & 58 & 52 \\
\hline
30   & 1.0 & 100 & 100 & 73 & 73 \\
30   & 1.5 &  97 & 100 & 74 & 73 \\
30   & 2.5 &  93 & 100 & 77 & 73 \\
\hline
24.5 & 1.0 & 100 & 100 & 89 & 89 \\
24.5 & 1.5 &  98 & 100 & 90 & 89 \\
24.5 & 2.5 &  95 & 100 & 92 & 89 \\

\enddata
\tablecomments{The 10\%-band filter is assumed to be rectangular from 9.975$\micron$-11.025$\micron$.  We also assume a flat SED and a site similar to the MMT's.  FWHM is measured in the altitude axis.  In the azimuth axis, FWHM is assumed to be diffraction-limited.}
\label{ELT_predictsnarrow}
\end{deluxetable}

\clearpage

\clearpage

\clearpage

\clearpage

\clearpage

\clearpage

\clearpage

\clearpage

\clearpage

\bibliographystyle{apj}
\bibliography{database}

\end{document}